\documentclass[aps,twocolumn,floats,floatfix,amssymb,nofootinbib,prd,superscriptaddress]{revtex4-1}
\usepackage{graphicx,amssymb,amsmath,amsthm,amsfonts,epsfig}
\usepackage[linktocpage,colorlinks=true,citecolor=OliveGreen,linkcolor=Maroon,urlcolor=Maroon]{hyperref}
\usepackage{appendix}
\usepackage{tensor}
\usepackage[normalem]{ulem}
\usepackage[dvipsnames, usenames]{xcolor}
\usepackage{nicematrix}
\usepackage{xr-hyper}
\definecolor{linkcolor}{rgb}{0.0,0.3,0.5}
\usepackage[all]{hypcap}
\usepackage[T1]{fontenc}
\usepackage[utf8]{inputenc}
\usepackage[usenames,dvipsnames]{xcolor}

\definecolor{romared}{RGB}{142,0,28}

\newcommand{\be}{\begin{equation}}
\newcommand{\ee}{\end{equation}}

\def\be{\begin{equation}}
\def\ee{\end{equation}}
\newcommand{\beq}{\begin{eqnarray}}
\newcommand{\eeq}{\end{eqnarray}}

\usepackage{makecell}
\usepackage{soul}

\newcolumntype{Y}{>{\centering\arraybackslash}X}

\definecolor{cornellGreen}{HTML}{6EB43F}

\begin{document}
\title{Phenomenology of ultralight bosons around compact objects:~in-medium suppression}
\author{Enrico Cannizzaro}
\affiliation{CENTRA, Departamento de F\'{\i}sica, Instituto Superior T\'ecnico -- IST, Universidade de Lisboa -- UL,
Avenida Rovisco Pais 1, 1049 Lisboa, Portugal}
\author{Thomas F.M.~Spieksma}
\affiliation{Niels Bohr International Academy, Niels Bohr Institute, Blegdamsvej 17, 2100 Copenhagen, Denmark}
\begin{abstract}
Mixing between ultralight bosons and the Standard Model photon may allow access to the hitherto invisible Universe. In the presence of plasma, photons are dressed with an effective mass which will influence the conversion between the two. We study this phenomenon, known as \emph{in-medium suppression}, in the context of black hole physics. We consider both axion--photon mixing around charged black holes and dark photon--photon mixing around neutral black holes. We find that the presence of plasma indeed influences the conversion rate, possibly quenching it altogether for large plasma densities, and discuss implications for superradiance and observational signatures.
\end{abstract}
\maketitle
%%%%%%%%%%%%%%%%%%%%%%%%%%%%%%%%%%%%%%%%%%%%%%%%%%%%
\section{Introduction}
%%%%%%%%%%%%%%%%%%%%%%%%%%%%%%%%%%%%%%%%%%%%%%%%%%%%
Black hole (BH) environments form prime territory to probe fundamental physics. An intriguing avenue concerns ultralight degrees of freedom, which have been proposed in various contexts such as the strong CP problem~\cite{Weinberg:1977ma,Wilczek:1977pj,Peccei:1977hh}, the dark matter problem~\cite{Preskill:1982cy, Abbott:1982af, Dine:1982ah, Jaeckel:2010ni, Feng:2010gw, Marsh:2015xka, Hui:2016ltb}, or in string theory (the so-called ``string axiverse'' scenario)~\cite{Arvanitaki:2010sy, Svrcek:2006yi,Mehta:2020kwu}. Rotating BHs are unstable against ultralight bosons, through a process called \emph{superradiance}~\cite{ZelDovich1971,ZelDovich1972,Brito:2015oca}. While the boson extracts rotational energy from the BH, a macroscopic ``cloud'' is formed, which can lead to striking observational signatures through e.g.~gravitational waves~\cite{Arvanitaki:2010sy,Arvanitaki:2014wva,Brito:2014wla,Brito:2015oca,Arvanitaki:2009fg,Brito:2017wnc,Brito:2017zvb,Hannuksela:2018izj,Baryakhtar:2017ngi,LIGOScientific:2021rnv,Tsukada:2018mbp, Palomba:2019vxe,Yuan:2022bem,Ng:2020jqd} or environmental effects~\cite{Baumann:2018vus, Baumann:2019ztm, Cardoso:2020hca, Baumann:2021fkf,Baumann:2022pkl,Tomaselli:2023ysb,Cole:2022fir,Zhang:2018kib,Zhang:2019eid,Berti:2019wnn,Tomaselli:2023ysb,Tomaselli:2024bdd,Tomaselli:2024dbw,Brito:2023pyl,Duque:2023seg,Boskovic:2024fga}.

Besides gravitational effects, interactions between spin-$0$ or spin-$1$ bosons and the Standard Model photon are realized through axionic couplings or vector ``portals'', respectively. In the former, the (pseudo)scalar is coupled to a two-photon vertex via a parity-violating term, while the vector portal is usually achieved through a kinetic mixing term~\cite{Holdom:2015kbf,Jaeckel:2010ni, Essig:2013lka, Fabbrichesi:2020wbt}. Such couplings have been searched for extensively both in laboratories~\cite{OSQAR:2015qdv, DellaValle:2015xxa, Ehret:2010mh, PhysRevD.42.1297, PhysRevLett.104.041301, Ouellet:2018beu, PhysRevLett.118.091801}
and astrophysical scenarios (see e.g.~\cite{Marsh:2015xka, Raffelt:1996wa, Caputo:2024oqc, Schlattl:1998fz, Vinyoles_2015, Redondo:2013wwa, Huang:2018lxq, Hook:2018iia, Leroy:2019ghm}). Even though the coupling strength might be weak, large densities of the ultralight field (achievable through, e.g.~superradiance), could lead to a new class of observational signatures~\cite{Chen:2022oad,Gan:2023swl}. In fact, recent work has shown that superradiant clouds can trigger powerful EM emission~\cite{Rosa:2017ury, Ikeda:2018nhb, Boskovic:2018lkj, Caputo:2021efm, Siemonsen:2022ivj, Spieksma:2023vwl, Xin:2024trp, Ferreira:2024ktd}. Additionally, axion-photon mixing is enhanced in the presence of strong EM fields. Such environments could be realized near neutron stars, whose strong magnetic field triggers a large conversion rate, making them ideal ``axion laboratories'' (see e.g.~\cite{Battye:2023oac, Huang:2018lxq, Hook:2018iia, Leroy:2019ghm, Witte:2020rvb, Witte:2021arp}). A striking phenomenology also occurs for charged BHs, which are unstable in the presence of axion-photon couplings~\cite{Boskovic:2018lkj}. This instability leads to the formation of axion ``hair'' in extremely short timescales and perturbative as well as full nonlinear solutions are now well studied~\cite{Boskovic:2018lkj,Burrage:2023zvk}.

Importantly, most studies ignore the impact of environmental plasmas, whose presence is ubiquitous around compact objects~\cite{Abramowicz:2011xu,Barausse:2014tra}, and which could strongly affect the resulting phenomenology. The key role of plasma is to provide the transverse degrees of freedom of a photon with an ``effective mass'', given by the plasma frequency:
\begin{equation}
\label{eq:plasmafreq}
   \omega_{\rm p}= \sqrt{\frac{n_{\rm e} e^2}{m_{\rm e}}}\approx \frac{10^{-12}}{\hbar}\sqrt{\frac{n_{\rm e}}{10^{-3}\text{cm}^{-3}}}\;\text{eV}\,,
\end{equation}
where $n_{\rm e}$ is the electronic density, while $e$ and $m_{\rm e}$ are the electron charge and mass, respectively. Whenever $\omega_{\rm p}\gg \mu$ (with $\mu$ the mass of the ultralight boson), the conversion is heavily suppressed, and stronger couplings are necessary for any significant production~\cite{PhysRevD.37.1237, Raffelt:1996wa, Mirizzi:2006zy, Redondo:2008aa, An:2013yfc}. This effect is referred to as \emph{in-medium suppression}, as the interaction between the boson and the photon is progressively weakened for denser plasmas. 

In this work, we study this phenomenon in the context of BH physics, by consistently taking into account the plasma dynamics on a BH spacetime. We focus on two relevant cases:~(i) axionic instabilities in Reissner-Nordstr\"{o}m spacetimes and (ii)~photon production from dark photon clouds around Schwarzschild BHs. While astrophysical BHs are expected to be uncharged due to different processes~\cite{Eardley:1975kp,Gibbons:1975kk,Cardoso:2016olt}, case (i) serves as a proxy to mimic axions around magnetized neutron stars. Moreover, the possibility that even a small charge, achievable via various mechanisms~\cite{Wald:1974np, Komissarov:2021vks, Cardoso:2016olt}, could lead to the formation of axionic hair is appealing. Case (ii) was partially considered in~\cite{Caputo:2021efm} using a plane-wave approximation, yet the full computation in curved spacetime was never performed. As highlighted in~\cite{Caputo:2021efm}, plasmas could play an important role in the evolution of superradiant instabilities, and prevent the interaction of the dark photon cloud with the Standard Model, especially for small dark photon masses and/or low couplings. 

This paper is organized as follows:~in Section~\ref{sec:theory} we review the setup of our model. In Section~\ref{sec:flat}, we discuss the impact of plasma on the mixing in flat spacetime. Then, in Section~\ref{sec:curved}, we turn our attention to BH spacetimes. Finally, we conclude in Section~\ref{sec:discussion}. We use $G=c=1$, unless otherwise stated and a ``mostly plus'' signature.

%%%%%%%%%%%%%%%%%%%%%%%%%%%%%%%%%%%%%%%%%%%%%%%%%%%%
\section{The theory}\label{sec:theory}
%%%%%%%%%%%%%%%%%%%%%%%%%%%%%%%%%%%%%%%%%%%%%%%%%%%%
We consider a generic Lagrangian involving a massive axion field $\Psi$ coupled to the EM field $A_{\mu}$. The EM field is also coupled to a dark photon (DP) field $A'_{\mu}$ via a kinetic mixing term $\sin\chi_0$, and is sourced by a cold, collisionless plasma. The DP sector is modelled in the so-called ``interaction basis''~\cite{Jaeckel:2012mjv, Siemonsen:2022yyf}. Other possible choices of basis are further discussed in Appendix \ref{appendix:DPbasis}.\footnote{When redefining the fields in the interaction basis, we neglect terms of the order $\left(\sin{\chi_0}\right)^{2}$~\cite{Jaeckel:2012mjv}. Note that typical values of the coupling are in the range $\sin{\chi_0} \lesssim \mathcal{O}(10^{-4})$~\cite{AxionLimits}.} We thus have:
\begin{equation}\label{eq:lagrangian}
    \begin{aligned}
        &\mathcal{L}=\frac{R}{16\pi}-\frac{1}{2}\nabla_{\mu}\Psi\nabla^{\mu}\Psi
-\frac{\mu_{\rm a}^{2}}{2}\Psi^{2}-\frac{k_{\rm a}}{2}\Psi\,{}^{*}\!F^{\mu\nu}F_{\mu\nu}+ j_\mu A^\mu\\&-\frac{1}{4}\left(F_{\mu\nu}F^{\mu\nu}+F'_{\mu\nu}F'^{ \mu\nu}\right)-\frac{\mu_{\gamma '}^2}{2} A'^\mu A'_\mu- \mu_{\gamma'}^2\sin\chi_0 A'_\mu A^\mu \\&+\mathcal{L}_{\rm m}\,,
\end{aligned}
\end{equation}
where $\mu_{\rm a}$ and $\mu_{\gamma'}$ are the axion and DP masses, $F_{\mu\nu}=\partial_\mu A_\nu-\partial_\nu A_\mu$ is the Maxwell tensor with an equivalent definition for $F'_{\mu\nu}$. Furthermore, ${}^{*}\!F^{\mu\nu}\equiv \frac{1}{2}\epsilon^{\mu\nu\rho\sigma}F_{\rho\sigma}$ is the dual Maxwell tensor and $R$ the scalar curvature. The plasma Lagrangian and current are given by $\mathcal{L}_{\rm m}$ and $j_\mu= e n_{\rm e} v_\mu$, respectively, with $v^\mu$ the electron's four velocity and $k_{\rm a}$ quantifies the strength of the axionic coupling. 

We model the plasma using an Einstein cluster setup~\cite{Einstein:1939ms,2012IJMPS..12..146G,Cardoso:2021wlq, Cardoso:2022whc, Feng:2022evy}, where plasma particles are assumed to be in circular orbits in all possible orientations around the BH. We refer to Appendix~\ref{app:Einsteincluster} for details on this setup, while we continue here to explain its main features. In the Einstein cluster, the plasma is described by the stress-energy tensor: 
\begin{equation}
\label{eq:plasmaSET}
    T^{\rm p}_{\mu\nu}=(\rho+P_{\rm t}) v_\mu v_\nu+P_{\rm t}(g_{\mu\nu}-r_\mu r_\nu)\,,
\end{equation}
where $\rho=n_{\rm e} m_{\rm e}$ is the energy density of the fluid, $P_{\rm t}$ the tangential pressure, $g_{\mu \nu}$ the metric of the underlying spacetime and $r^\mu$ a unit vector in the radial direction.

From the Lagrangian~\eqref{eq:lagrangian}, we can infer the equations of motion for the scalar, EM, DP and gravitational fields: 
\begin{equation}\label{eq:evoleqns}
\begin{aligned}
\left(\nabla^\mu \nabla_\mu-\mu_{\rm a}^2\right) \Psi &= \frac{k_{\mathrm{a}}}{2}\,{ }^*\!F^{\mu \nu} F_{\mu \nu}\,,\\
\nabla_\nu F^{\mu \nu} &=j^{\mu} -2 k_{\mathrm{a}}{ }^*\!F^{\mu \nu} \nabla_\nu \Psi-\sin\chi_0 \mu_{\gamma'}^2 A' {}^{\mu}\,,\\
\nabla_\nu F'^{\mu \nu} &=-\mu_{\gamma'}^2 A'^\mu-\sin\chi_0 \mu_{\gamma'}^2 A^{\mu}\,,\\
R_{\mu\nu}-\frac{1}{2}g_{\mu\nu}R&= 8 \pi \Big(T^{\rm \Psi}_{\mu\nu}+T^{\rm EM}_{\mu\nu}+T^{\rm DP}_{\mu\nu}+T^{\rm p}_{\mu\nu} \Big)\,,
\end{aligned}
\end{equation}
where we introduced the Ricci tensor $R_{\mu\nu}$ and the stress-energy tensor of the axionic, EM and DP sector. 

To close the system, we need the continuity and momentum equations, which we infer from the conservation of the current and the stress-energy tensors:
\begin{equation}
\label{eq:momentum_continuity}
\nabla^\nu T^{\rm p}_{\mu\nu}=e n_{\rm e} F_{\mu\nu}\,, \quad \nabla_\mu (n_{\rm e} v^\mu)=0\,.
\end{equation}
As we choose to model the DP sector in the interaction basis, the hidden field $A'_\mu$ is \emph{sterile}, meaning that it does not couple to the plasma directly.
%%%%%%%%%%%%%%%%%%%%%%%%%%%%%%%%%%%%%%%%%%%%%%%%%%%%
\section{Plasma suppression in flat spacetime}
\label{sec:flat}
%%%%%%%%%%%%%%%%%%%%%%%%%%%%%%%%%%%%%%%%%%%%%%%%%%%%
Before studying BH spacetimes, we consider the impact of plasma on the mixing in flat spacetime. This allows us to isolate and explain some of the key dynamics. In flat space, the Einstein cluster model simply reduces to a plasma at rest, with a vanishing tangential pressure. We can thus treat the plasma as pressureless dust, with a momentum equation given by
\begin{equation}\label{eq:momentumflat}
    v^\nu \partial_\nu v^\mu=\frac{e}{m_{\rm e}}F^{\mu\nu}v_\nu\,.
\end{equation}

In presence of an external magnetic field, there is mutual conversion between the axion and the propagating modes of the photon~\cite{PhysRevLett.51.1415,PhysRevD.37.1237, Raffelt:1996wa, Mirizzi:2006zy}, which parallels the mixing of neutrinos. On the other hand, dark photon-photon mixing can arise even in the absence of background EM fields.

First, we consider the axionic case, setting $\sin\chi_0=0$. As a background configuration, we take a static, homogeneous electron-ion plasma in a constant magnetic field along the $\hat{y}$ direction. We then adopt linear perturbation theory to study the propagation of plane waves. As detailed in~\cite{PhysRevD.37.1237}, the conversion between axions and photons requires a change in the azimuthal angular momentum. A longitudinal magnetic field would preserve azimuthal symmetry, thus stopping any conversion. Hence, we consider propagation of waves along the $\hat{z}$-axis without loss of generality. 

We denote the components of the EM potential as parallel ($A_y=A_\parallel$) and perpendicular ($A_x=A_\perp$) to the background magnetic field. As for the axionic sector, we consider a vanishing axion background $\Psi = 0$, and denote its perturbation by $\psi$. From parity considerations, one can readily see that the perpendicular component of the EM field decouples from the axion. Indeed, the two photon states $A_\perp$ and $A_\parallel$ are even and odd, respectively, under parity in the $y-z$ plane, while the axion plane wave state is odd. Thus, only the parallel component of the EM field mixes with the axion~\cite{PhysRevD.37.1237}. Considering only the dynamics along this direction, solving the momentum equation~\eqref{eq:momentumflat} yields
\begin{equation}
    v_\parallel=-\frac{e}{m_{\rm e}}A_\parallel\,,
\end{equation}
so that the current can be expressed in terms of the EM field. One is then left with a coupled system involving the parallel component of Maxwell equations and the Klein-Gordon equation. In the frequency-domain and assuming for simplicity relativistic axions ($\omega\gg \mu_{\rm a}$), it can be expressed as\footnote{Note that, in the following we will neglect the impact of Faraday rotation. Although it can be important for polarization effects, it does not affect the relevant subject here, which is the conversion probability between axions and photons~\cite{Mirizzi:2006zy}.}
\begin{equation}\label{eq:axion-photon-mixing-flat}
    (\omega- i\partial_z+\mathcal{M}_{\mathrm{a}\gamma})\begin{bmatrix}
    A_\parallel \\
    \psi
    \end{bmatrix} =0\,,
\end{equation}
where we introduced the ``axion-photon mixing matrix'':
\begin{equation}
\label{eq:Edispersion}  
\mathcal{M}_{\mathrm{a}\gamma }=
    \begin{bmatrix}
    -\omega_{\rm p}^2/2 \omega & B_y k_{\rm a}  \\
    B_y k_{\rm a} & -\mu_{\rm a}^2/2\omega 
    \end{bmatrix}\,.
\end{equation}
Here, the off-diagonal terms couple the two fields, and are thus responsible for the mixing. Clearly, in the case $k_{\rm a}=0$ or $B_y=0$, i.e., when the matrix is diagonal, the two fields are decoupled, and Eq.~\eqref{eq:axion-photon-mixing-flat} simply returns the dispersion relation of the photon and axion. To simplify the dynamics, we can perform a field redefinition that diagonalizes the matrix~\eqref{eq:Edispersion}.\footnote{This is similar to adopting a field redefiniton in Reissner-Nordstr\"{o}m spacetimes to decouple the gravito-electromagnetic perturbations in terms of two master functions (see e.g.~\cite{Chandrasekhar:1985kt, Pani:2013wsa, Berti:2005eb}).}
To do so, it is sufficient to perform a rotation in the field basis by the rotation angle:
\begin{equation}\label{eq:angle-axion-photon}
    \theta= \frac{1}{2}\arctan{\left(\frac{4 \omega B_y k_{\rm a}}{-\omega_{\rm p}^2+\mu_{\rm a}^2} \right)}\,.
\end{equation}
Note that this angle is proportional to the off-diagonal terms, and thus quantifies the coupling between the modes: indeed, the axion-photon conversion rate is proportional to $P(a \rightarrow \gamma)\propto \rm{sin}^2 (2 \theta)$~\cite{Mirizzi:2006zy, PhysRevD.37.1237}. In the limits $B_y \rightarrow 0$ or $k_{\rm a} \rightarrow 0$, the probability naturally goes to zero and no conversion is possible. However, even in the presence of magnetic fields and couplings, a large plasma frequency, i.e., $\omega_{\rm p}\gg \mu_{\rm a}, k_{\rm a} B_y$, kinematically disfavours the conversion from axions to photons, as the mixing angle drops to zero. This quenching is referred to as \emph{in-medium suppression}. Moreover, when $\mu=\omega_{\rm p}$, the conversion probability is maximized, which is termed a \textit{resonant conversion} between the two states.

In the DP case ($k_{\rm a}=0$), the above procedure yields the dark photon-photon mixing matrix~\cite{An:2023mvf}:
\begin{equation}
\label{eq:DPEdispersion}  
\mathcal{M}_{\gamma \gamma' }=\frac{1}{2\omega}
    \begin{bmatrix}
    -\omega_{\rm p}^2 & \sin{\chi_0}\, \mu_{\gamma'}^2  \\
    \sin{\chi_0}\,\mu_{\gamma'}^2 & -\mu_{\gamma'}^2
    \end{bmatrix}\,.
\end{equation}
Similar conclusions applies as in the axionic case: in the presence of a sufficiently dense plasma, i.e, $\omega_{\rm p}\gg \mu_{\gamma'}$, the in-medium conversion angle goes to zero, suppressing the conversion. When $\mu_{\gamma'}=\omega_{\rm p}$ instead, a resonant conversion is triggered.

%%%%%%%%%%%%%%%%%%%%%%%%
\section{Plasma suppression in curved spacetime}\label{sec:curved}
%%%%%%%%%%%%%%%%%%%%%%%%
\subsection{Charged black holes: axionic instabilities}\label{sec:instability_RN}
%%%%%%%%%%%%%%%%%%%%%%%%
We now turn to a fully relativistic setup, and analyze axionic instabilities around charged compact objects using BH perturbation theory. We ignore the backreaction of the plasma in the Einstein and Maxwell background equations~\cite{Cannizzaro:2024yee}, since source terms are suppressed by the large charge-to-mass ratio of the electron and the low energy density of typical astrophysical environments. Moreover, we consider the presence of an oppositely charged component in the plasma---ions---inducing a current with the opposite sign in the Maxwell equations, neutralizing the plasma background~\cite{Cannizzaro:2020uap,Cannizzaro:2021zbp,Cannizzaro:2023ltu,Spieksma:2023vwl}. Assuming a spherically symmetric spacetime, the background geometry is then the standard Reissner-Nordstr\"{o}m (RN) solution:
\begin{equation}
\label{eq:RN_Sol}
\begin{aligned}
\mathrm{d}s^2 & =-f\mathrm{d} t^2+f^{-1} \mathrm{d} r^2+r^2\mathrm{d}\Omega^{2}\,, \\
&\text{with}\quad f= 1-\frac{2 M}{r}+\frac{Q^2}{r^2}\,, 
\end{aligned}
\end{equation}
where $M$ and $Q$ are BH mass and charge respectively, and the background EM field is given by $A_\mu=(Q/r,0,0,0)$. 

We model the plasma as non-relativistic and consider a macroscopic plasma effective mass, $M\omega_{\rm p}\gtrsim 1$. As shown in~\cite{Cannizzaro:2024yee} however, any significant value of the BH charge induces a relativistic plasma motion, which partially suppresses the photon's effective mass via relativistic plasma transparency~\cite{KawDawson1970, Cardoso:2020nst, Cannizzaro:2024yee}. This is realized through the Lorentz factor $\gamma$ in the electron's relativistic mass $m_{\rm e}\rightarrow \gamma m_{\rm e}$, which enters the denominator of the plasma frequency~\eqref{eq:plasmafreq}. Yet, a number of effects such as screening of the BH charge~\cite{Feng:2022evy} and magnetic field pressure, can contrast the relativistic motion. Furthermore, this (non-relativistic) model is intended as a proxy to study compact objects embedded in EM fields in more realistic scenarios. The most notable example is magnetized neutron stars, where surrounding plasma is expected to dress the photon with a large plasma frequency (see e.g.~\cite{Leroy:2019ghm, Witte:2020rvb, McDonald:2023shx}). While further steps are required to model realistic astrophysical scenarios, this is the first analysis that incorporates plasma effects on axion-photon conversion in a consistent curved spacetime model using BH perturbation theory. More details are provided in Appendix~\ref{app:Einsteincluster}.

The presence of background EM fields in electrovacuum charged BHs induces axionic instabilities, leading to new ``hairy'' BH solutions~\cite{Ikeda:2018nhb}. In this section, we scrutinize whether axion-photon interactions also arise in more astrophysically relevant scenarios, where BHs are surrounded by plasmic environments. In Appendix~\ref{appendix:flatinstab}, we show a flat spacetime analysis in a simplified setup, which makes the problem analytically tractable, yet still elucidates some of the phenomena.

To study the system in a RN background, we linearize the field equations~\eqref{eq:evoleqns} and perform a multipolar decomposition of the fields. As the geometry is spherically symmetric, perturbations can be recasted in two decoupled sectors, axial and polar, depending on their behaviour under parity transformations~\cite{Regge:1957td,Zerilli:1970wzz,Zerilli:1970se,Zerilli:1974ai}. Due to its pseudo-scalar nature, the axion only couples with the axial sector of the system, which is thus the only sector of interest. The polar sector instead, is described by the standard gravito-EM perturbations (in the presence of plasma).
Details on the perturbation scheme are reported in Appendix~\ref{appendix:multipolarexp}. In the axial sector, perturbations of the fluid can be solved analytically. Indeed, the pressure and density perturbation are scalar quantities with a polar symmetry, and thus vanish identically, while the axial fluid velocity--$v_4$--can be related to the axial EM mode--$u_4$--via the linear relation 
\begin{equation}
\label{eq:momentumcons}
v_{4}=-\frac{e}{m_{\rm e}}u_{4}\,.
\end{equation}
The system is then described by three variables: the gravitational Moncrief-like master variable $\Psi_{\rm m}$, the EM axial degree of freedom $u_{4}$ and the axion multipole field $\psi$. These three functions obey a set of coupled, second order partial differential equations:
\begin{equation}
\label{eq:wavelike-eqn}
\begin{aligned}
\hat{\mathcal{L}}\Psi_{\rm m} &=\Bigg(\frac{4 Q^4}{r^6}+ \frac{Q^2(-14 M + r (4+\lambda))}{r^5} \\&+ \left(1-\frac{2M}{r}\right)\left[\frac{\lambda}{r^{2}}-\frac{6M}{r^{3}}
\right]\Bigg)\Psi_{\rm m}-\frac{8 Q f}{r^3 \lambda} u_{4}\,, \\
\hat{\mathcal{L}}u_{4}&=f\left(\omega_{\rm p}^2 + \frac{\lambda}{r^{2}} + \frac{4 Q^2}{r^4}\right) u_{4}-\frac{(\ell-1)\lambda(\ell+2) Q f}{2 r^3}\Psi_{\rm m} \\&
+\frac{2 \lambda Q f k_{\rm a}}{r^3}\psi\,,
 \\
\hat{\mathcal{L}}\psi&=f\Bigg(\frac{-2 Q^2+r(2M+r \lambda +r^3 \mu_{\rm a}^2)}{r^4}\Bigg) \psi +\frac{2 Q  f k_{\rm a} }{r^2} u_{4}\,,
\end{aligned}
\end{equation}
where $\lambda=\ell(\ell+1)$, $\hat{\mathcal{L}} = \partial^{2}/\partial r_*^{2} - \partial^{2}/\partial t^{2}$ and the tortoise coordinate is defined as $\mathrm{d}r_*/\mathrm{d}r = f^{-1}$. Note that in the limit $Q \rightarrow 0$, the axion decouples from the system. In other words, in the absence of background EM fields, no mixing with the photon is possible.

In the following, we will solve these equations using the numerical routine presented in~\cite{Cannizzaro:2024yee}. Analogous to that work, we initialize each sector with a Gaussian, choosing its amplitude, frequency and width to be $A = 1$, $M\Omega_0 = 0.4$, $\sigma = 4.0M$. We initialize at $r_0 = 20M$. 
\begin{figure}
    \centering
    \includegraphics[width = \linewidth]{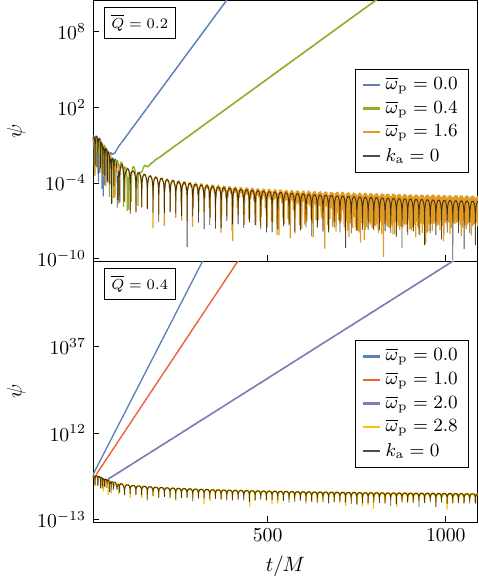}
    \caption{Evolution of the axion sector $\psi$ for $\overline{Q} = 0.2$ (\emph{top panel}) and $\overline{Q} = 0.4$ (\emph{bottom panel}) and various choices of the plasma frequency (quantities with a bar are dimensionless, e.g.~$\overline{Q} = Q/M$). We initialize with a Gaussian in all sectors (gravitational, EM and axion) and choose $k_{\rm a} = 20$, $\ell = 2$, $\mu_{\rm a} M = 0.2$, while we extract at $r_{\rm ex} = 30M$. In absence of plasma ($\overline{\omega}_{\rm p} = 0$), the instability rates follow Eq.~(47) in~\cite{Boskovic:2018lkj}. For fixed coupling, the critical plasma frequency to quench the instability scales as $\overline{\omega}^{\rm crit}_{\rm p} \propto \overline{Q}^{3/2}$. When the photon production is heavily suppressed (large $\omega_{\rm p}$), the behaviour of the axionic sector limits toward that of a free scalar (black line).}
    \label{fig:axion_hair}
\end{figure}

Figure~\ref{fig:axion_hair} shows the evolution of the axion field for different values of the plasma frequency. The two panels corresponds to two different values of the BH charge $\overline{Q}=0.2, 0.4$, where we will denote dimensionless quantities with an overhead bar. As evident from the figure, the system is unstable to axionic perturbations for $\overline{\omega}_{\rm p}=0$, provided that the BH charge or the axionic coupling is sufficiently high. We find a good agreement with the instability condition found in~\cite{Ikeda:2018nhb} in this limit. Nevertheless, increasing $\omega_{\rm p}$ progressively weakens the instability, quenching it altogether after a critical value. This is similar to the flat spacetime analysis reported in Appendix~\ref{appendix:flatinstab}: large values of $\omega_{\rm p}$ tend to stabilize the system, and higher values of $Q$ or $k_{\rm a}$ are needed to restore the instability. Indeed, as shown in the bottom panel, for a higher value of $Q$, a larger $\omega_{\rm p}$ is needed to quench the growth. More specifically, at fixed couplings $k_{\rm a}$ the critical plasma frequency necessary to quench the instability scales as $\overline{\omega}^{\rm crit}_{\rm p} \propto \overline{Q}^{3/2}$.

Finally, whenever the instability is quenched, the axion behaves as a free scalar field in a RN geometry: this is because plasma effectively decouples it from the photon. We show this in Figure~\ref{fig:axion_hair}, where the axion field in the regime $k_{\rm a} \neq 0, \omega_{\rm p} \gg \mu_{\rm a}$ coincides with that of a axion field with $k_{\rm a}=0$, denoted by the black line. This is in agreement with the discussion in the previous section and Appendix~\ref{appendix:flatinstab} (see Figure~\ref{fig:flatmodes}).
%%%%%%%%%%%%%%%%%%%%%%%%%%%%%%%%%%%%%%%%%%%%%%%%%%%%%%%%%%%%%%
\subsection{Dark photon clouds:~photon production}
\label{sec:darkphotons}
%%%%%%%%%%%%%%%%%%%%%%%%%%%%%%%%%%%%%%%%%%%%%%%%%%%%%%%%%%%%%
Consider then dark photon-photon mixing in a BH background. Since no background EM fields are necessary for the mixing to occur, we stick to a Schwarzschild spacetime, with the line element given by~\eqref{eq:RN_Sol} with $Q=0$, so that $f=1-2M/r$. Similar to before, we consider a non-relativistic background plasma. 

Setting $k_{\rm a} = 0$, we linearize the field equations~\eqref{eq:evoleqns} and perform a multipolar expansion using the ansatz reported in Appendix~\ref{appendix:multipolarexp}. Unlike the axionic case, DP perturbations appear both in the axial and polar sector, and couple with the EM one. For simplicity, we will stick to the axial sector in this work, as fluid perturbations can be solved for analytically.

In the absence of a background charge, gravitational perturbations decouple and thus with respect to Eq.~\eqref{eq:wavelike-eqn}, the system is now governed by only two master functions:~the axial EM mode $u_{4}$ and the DP mode $u'_{4}$. They satisfy the following system of coupled partial differential equations:
\begin{equation}
\label{eq:wavelike-eqn_DP}
\begin{aligned}
\hat{\mathcal{L}}u_{4}&=f\left(\omega_{\rm p}^2 + \frac{\lambda}{r^{2}}\right) u_{4}+f \mu_{\gamma'}^2 \sin{\chi_0}\,u'_{4}\,,
 \\
\hat{\mathcal{L}}u'_{4}&=f\left(\mu_{\gamma'}^2 + \frac{\lambda}{r^{2}}\right) u'_{4}+f \mu_{\gamma'}^2 \sin{\chi_0}\,u_4\,,
\end{aligned}
\end{equation}
where $\mathcal{L}$ and $\lambda$ were defined below~\eqref{eq:wavelike-eqn}. Clearly, from~\eqref{eq:wavelike-eqn_DP}, the mixing between the modes is proportional to the vacuum mixing angle $\sin{\chi_0}$ as well as the DP mass, consistent with the flat spacetime analysis~\eqref{eq:DPEdispersion}.  Note that, in the current analysis the field equations~\eqref{eq:wavelike-eqn_DP} are computed in the interaction basis. For completeness, in Appendix~\ref{appendix:DPbasis}, we also report the equivalent of~\eqref{eq:wavelike-eqn_DP} in the mass basis. We now evolve the equations of motion in time, starting with initial data \emph{only} in the DP sector.
\begin{figure}
    \centering
    \includegraphics[width = \linewidth]{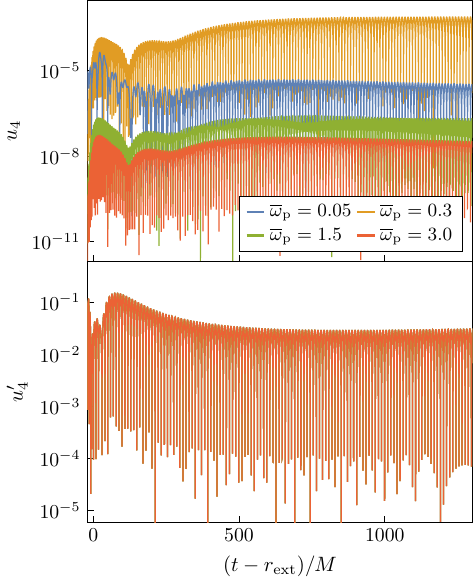}
    \caption{Evolution of the EM (\emph{top panel}) and DP (\emph{bottom panel}) field for various choices of the plasma frequency. We initialize purely in the DP sector, and take $\sin{\chi_0} = 0.0001$, $\mu_{\gamma'}M = 0.3$ and $\ell = 1$. In the top (bottom) panel, we extract at $r_{\rm ex} = 400\,(50)\,M$. The DP field settles on a QBS, whose frequency we have checked explicitly: $\overline{\omega}_{\rm R} = 0.29564$ (time-domain) vs. $\overline{\omega}_{\rm R} = 0.29598$ (frequency-domain).}
    \label{fig:DarkPhotonPhoton}
\end{figure}

The results are shown in Figure~\ref{fig:DarkPhotonPhoton}. In the bottom panel, we show the DP field which is barely affected by the mixing due to the small coupling. The evolution of the initial data thus simply entails outward travelling waves as well as the formation of a quasi-bound state (QBS) near the BH. The latter is the component we show, as we extract the field just after the peak density of the bound state, and we checked explicitly its frequency corresponds to the QBS one (see the caption). 

In the top panel instead, we show the EM field, which is clearly affected by the plasma. Its production is peaked at $\omega_{\rm p}=\mu_{\gamma'}=0.3/M$, where the conversion probability is expected to be resonantly enhanced. When increasing the plasma density, i.e., $\omega_{\rm p}\gg \mu_{\gamma'}$, the in-medium suppression acts and the EM field decays. In this regime, the two fields decouple more and more, such that less photons are produced. The rate at which this happens as a function of the plasma frequency, has been studied analytically in flat spacetime ~\cite{Dubovsky:2015cca,Caputo:2021efm}. It was found that for large plasma frequencies, the suppression factor should be $\propto \mu_{\gamma'}^2/\omega_{\rm p}^{2}$.
\begin{figure}
    \centering
    \includegraphics[width = \linewidth]{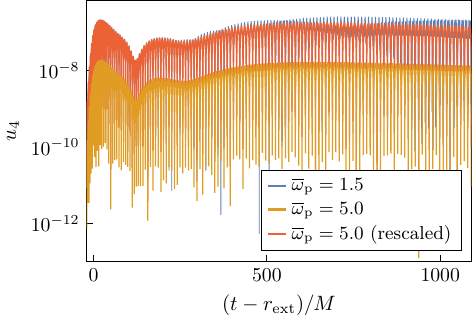}
    \caption{Similar setup as Figure~\ref{fig:DarkPhotonPhoton} for $\overline{\omega}_{\rm p} = 1.5, 5.0$. The suppression goes as $\propto \omega_{\rm p}^{-2}$, which is the rescaling applied to the red curve.}
    \label{fig:DP_scaling}
\end{figure}

We verify this scaling numerically in a BH background in Figure~\ref{fig:DP_scaling}, which shows the EM field evolution for a fixed boson mass ($\mu_{\gamma'}=0.3/M$) and $\bar{\omega}_{\rm p}=1.5, 5$. Both cases correspond to a regime where $\omega_{\rm p}\gg \mu_{\gamma'}$, such that the scaling in~\cite{Dubovsky:2015cca, Caputo:2021efm} should hold. The two curves (blue and red) coincide incredibly well, especially in the earlier part of the signal. In this regime, the signal consist purely of relativistic modes, with $\omega \gg \omega_{\rm p}$, such that dispersion effects are negligible. At later times, the signal is ``contaminated'' by non-relativistic modes with $\omega \gtrsim \omega_{\rm p}$, and the scaling is slightly spoiled by dispersion effects.

Confirming the predicted scaling from~\cite{Dubovsky:2015cca, Caputo:2021efm} allows us to estimate the strength of the electric field generated from a superradiant DP cloud. In particular, we find the ratio between the EM and DP field amplitudes to be
\begin{equation}
\frac{A_{\gamma}}{A_{\gamma'}} = 10^{-17}\left(\frac{\sin{\chi_0}}{10^{-7}}\right)\left(\frac{\mu_{\gamma'}}{10^{-15}\,\mathrm{eV}}\right)^{2}\left(\frac{10^{-10}\,\mathrm{eV}}{\omega_{\rm p}}\right)^{2}\,.
\end{equation}
Considering the scenario of a fully grown cloud~\cite{East:2017ovw,East:2017mrj} (such that the mass of the cloud is $10$\% of the BH mass), we can estimate the strength of the observable electric field, which will depend on the ``in-medium suppression factor'' ($\mu_{\gamma'}/\omega_{\rm p}$) as
\begin{equation}\label{eq:electricfield_DPgenerated}
E_0 \simeq 6.3 \left(\frac{\sin{\chi_0}}{10^{-7}}\right) \left(\frac{\mu_{\gamma'}/\omega_{\rm p}}{10^{-7}}\right) \left(\frac{\alpha}{0.2}\right)^3 \left(\frac{10^6 M_{\odot}}{M}\right)\,\frac{\mathrm{V}}{\mathrm{m}}\,,
\end{equation}
where $\alpha \equiv \mu_{\gamma'} M$ is the ``gravitational coupling'', essential to the superradiance process. Note that, fixing $\alpha$ and $M$ implies a value for the boson mass, e.g.~for $\alpha = 0.2$ and $M = 10^{6}M_{\odot}$, the mass of the boson is $\mu_{\gamma'} = 2.6 \times 10^{-17}\,\mathrm{eV}$. To convert from an electronic density to the plasma frequency, one can then make use of Eq.~\eqref{eq:plasmafreq}.
%%%%%%%%%%%%%%%%%%%%%%%%%%%%%%%%%%%%%%%%%%%%%%%%%%%%
\section{Discussion}
\label{sec:discussion}
%%%%%%%%%%%%%%%%%%%%%%%%%%%%%%%%%%%%%%%%%%%%%%%%%%%
Most conjectured ultralight particles are expected to couple weakly to the Standard Model, making a possible detection challenging. However, in regions of strong gravity, such as around compact objects, they could play a central role. Consequently, there has been a huge effort from the community to characterize possible observational signatures, many of which involve EM radiation. 

Most studies however, ignore the impact of plasma completely, or make use of simple flat spacetime approximations. As the majority of astrophysical compact objects is expected to be surrounded by plasma, e.g.~in the form of accretion disks, a more detailed analysis is imperative. In this work, we introduce a consistent, fully relativistic framework to study the impact of plasma on the dynamics of ultralight bosons around compact objects. Plasma is found to play a pivotal role when bosons are coupled to the Maxwell sector. Two examples are studied explicitly: axion-photon mixing around charged BHs and dark photon-photon mixing around neutral BHs. Both scenarios yield similar conclusions:~the presence of plasma strongly impacts the conversion rate and suppresses it in realistic regimes. This effect, termed \emph{in-medium suppression}, could have an impact on the observational signatures from such systems and thus on constraints that have been set on the coupling constants. In addition, plasma thus makes the evolution of superradiant systems more robust by preventing the conversion of bosons into photons. This strengthens the possibility to detect superradiant clouds, e.g.~through binary systems~\cite{Baumann:2018vus, Baumann:2019ztm, Cardoso:2020hca, Baumann:2021fkf,Baumann:2022pkl,Tomaselli:2023ysb,Cole:2022fir,Zhang:2018kib,Zhang:2019eid,Berti:2019wnn,Tomaselli:2023ysb,Tomaselli:2024bdd,Tomaselli:2024dbw,Brito:2023pyl,Duque:2023seg,Boskovic:2024fga}. 

As an example, consider typical densities of the interstellar medium $n_{\rm e}\!\sim\! 10^{-3}\!-\!10\,\mathrm{cm}^{-3}$, which lead to a plasma frequency of $\omega_{\rm p}\!\sim\! 10^{-12}\!-\!10^{-10} \textrm{eV}$~\cite{Cordes:2002wz,Conlon:2017hhi, Pani:2013hpa}. Such densities are already sufficient to suppress mixing in most of the superradiant mass range of ultralight bosons $\mu\!\sim\! 10^{-20}\!-\!10^{-10} \textrm{eV}$. Moreover, densities of accretion disks are typically many orders of magnitude larger than that of the interstellar medium~\cite{Abramowicz2013, Barausse:2014tra}, resulting in an even stronger suppression.

\begin{figure}
    \centering
    \includegraphics[width = \linewidth]{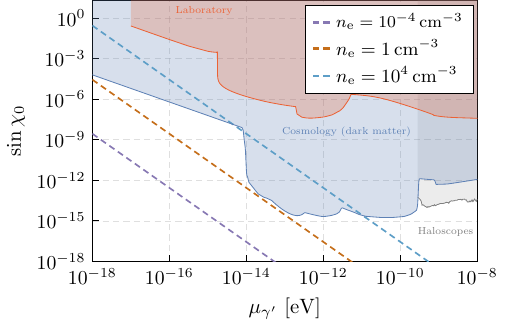}
    \caption{Kinetic mixing for varying DP mass. Below the dashed lines, in-medium suppression is efficient and cannot be removed by non-linear effects. Typical densities of the interstellar medium $\left(n_{\rm e}\!\sim\! 10^{-3}\!-\!10\,\mathrm{cm}^{-3}\right)$ are sufficient to suppress the mixing in most of the unconstrained regions of the parameter space. Shaded regions show constraints from cosmological nature (blue), experiments on earth (red) or haloscopes (gray), see~\cite{ Caputo:2021eaa,AxionLimits} and references therein.}
    \label{fig:DP_bounds}
\end{figure}

A natural extension of our work would be to generalize our framework to the case of magnetized neutron stars. Recent work showed the axion phenomenology in that scenario can be extremely rich, in particular in small localized regions  of the magnetosphere called ``polar caps'', where EM fields are large enough to produce an enormous quantity of axions~\cite{Prabhu:2021zve, Noordhuis:2022ljw}. Such axions may stream away and resonantly convert into photons, generating broadband radio fluxes. Alternatively, they could bound gravitationally to the star, forming axion clouds with extreme densities~\cite{Noordhuis:2023wid}. Such clouds generate an axion-induced electric field leading to striking EM observables, like periodic nulling in the pulsar emission or complementary radio emissions~\cite{caputo:2023cpv}. The electric field is proportional to the in-medium suppression factor ($\mu_{\rm a}/\omega_{\rm p}$), analogous to our discussion on dark photons. Applying our framework to such systems requires a careful modelling of the background system, in particular the magnetic field geometry and the magnetosphere phenomenology. In addition, there will be mixing between $\ell$ modes in the presence of magnetic fields (see~\cite{Brito:2014nja, Day:2019bbh}).

Another interesting direction are nonlinear photon-plasma interactions in the context of DP superradiance. In~\cite{Caputo:2021efm}, it is discussed how in-medium suppression could be avoided for low electron densities in a compelling region of the parameter space, i.e., $m_{\gamma'}\gtrsim 10^{-16}\,\rm{eV}$ and $\sin{\chi_0} \gtrsim 10^{-8}\!-\!10^{-7}$, due to relativistic transparency effects induced by the DP cloud. A similar assumption was made in~\cite{Siemonsen:2022ivj}, where it is shown how superradiant DP field may generate a pair plasma via the Schwinger mechanism. Due to the large value of the ambient electric field when the plasma is formed, the plasma frequency is assumed to vanish. 

Using~\eqref{eq:electricfield_DPgenerated}, we can roughly estimate the required values for the plasma frequency to be quenched, as the electric field must be high enough to induce nonlinear plasma dynamics, i.e., $E_{\rm NL} \approx m_{\rm e}\omega_{\rm p}/e$~\cite{Cardoso:2020nst,Cannizzaro:2023ltu}. Figure~\ref{fig:DP_bounds} shows a lower bound on the mixing value for which non-linear effects can remove the in-medium suppression. As evident from the figure, typical densities of the interstellar medium are sufficient to suppress the mixing in most of the unconstrained regions of the parameter space. For more dense plasmic environments such as accretion disks, the necessary boson mass for nonlinear effects may fall outside of the allowed superradiant range entirely. Including nonlinearities might allow for a more precise investigation of this transparency effect as well as a wide array of other, potentially relevant effects (see e.g.~\cite{Cannizzaro:2023ltu}). 
%%%%%%%%%%%%%%%%%%%%%%%%
\begin{acknowledgments}
%%%%%%%%%%%%%%%%%%%%%%%%
%
We thank Andrea Caputo and Yifan Chen for a critical reading of the manuscript and Richard Brito and David Marsch for useful comments. This work was supported by the VILLUM Foundation (grant no.\ VIL37766) and the DNRF Chair program (grant no.\ DNRF162) by the Danish National Research Foundation. We acknowledge financial support provided under the European Union’s H2020 ERC Advanced Grant “Black holes: gravitational engines of discovery” grant agreement no.\ Gravitas–101052587. Views and opinions expressed are however those of the author only and do not necessarily reflect those of the European Union or the European Research Council. Neither the European Union nor the granting authority can be held responsible for them. The authors have received funding from the European Union's Horizon 2020 research and innovation programme under the Marie Sk{\l}odowska-Curie grant agreement No. 101007855 and No. 101131233.
%%%%%%%%%%%%%%%%%%%%%%%%
\end{acknowledgments}
%%%%%%%%%%%%%%%%%%%%%%%%

\appendix
%%%%%%%%%%%%%%%%%%%%%%%%
\section{Einstein cluster}\label{app:Einsteincluster}
%%%%%%%%%%%%%%%%%%%%%%%%
Following earlier work~\cite{Cannizzaro:2024yee}, we model the plasma using a so-called ``Einstein cluster''~\cite{Einstein:1939ms,2012IJMPS..12..146G,Cardoso:2021wlq,Cardoso:2022whc,Feng:2022evy,Cannizzaro:2024yee}. In this model, the plasma consists of many charged particles orbiting in circular motion in all possible directions. Performing an angular average over those orbits then reduces to having an anisotropic, static fluid with a non-vanishing tangential pressure. It is described by the stress-energy tensor: 
\begin{equation}\label{eq:stress_energy_appendix}
T^{\rm p}_{\mu\nu}=(\rho+P_{\rm t}) v_\mu v_\nu+P_{\rm t}(g_{\mu\nu}-r_\mu r_\nu)\,,
\end{equation}
where $\rho=n_{\rm e} m_{\rm e}$ is the energy density of the fluid, $v^\mu$ the four-velocity, $P_{\rm t}$ the tangential pressure, $g_{\mu \nu}$ the spacetime metric and $r^\mu$ a radial unit vector. 

As detailed in the main text, we treat the plasma as non-relativistic. This assumption implies that the tangential pressure, which incorporates the electron angular velocity, is way smaller than the energy density, i.e., $P_{\rm t}\ll\rho$. Then,~\eqref{eq:stress_energy_appendix} reduces to
\begin{equation} 
T^{\rm p}_{\mu\nu}\approx \rho v_\mu v_\nu+P_{\rm t}(g_{\mu\nu}-r_\mu r_\nu)\,,
\end{equation}
while the momentum equation~\eqref{eq:momentum_continuity} takes on the form of the non-relativistic Euler equation. Solving this equation yields the tangential pressure of the background, which reads
\begin{equation}
\label{eq:tangentialpressure}
    P_{\rm t}=-\frac{n_{\rm e}(e Q r\sqrt{f}+(Q^2-Mr)m_{\rm e})}{Q^2-3Mr+2r^2}\,,
\end{equation}
where we take $Q=0$ in Section~\ref{sec:darkphotons}. Note that a similar calculation can be done for relativistic particle motion. This solution, as well as the resulting phenomenology, are discussed in~\cite{Cannizzaro:2024yee} (see in particular Appendix S.2). The Einstein cluster thus provides a simplified, but realistic description of the motion of matter around BHs.

%%%%%%%%%%%%%%%%%%%%%%%%
\section{Perturbation equations}\label{appendix:multipolarexp}
%%%%%%%%%%%%%%%%%%%%%%%%
In this appendix, we provide details on the perturbation scheme. From the background solution, we linearize the field
equations. We denote quantities by $X = X^{(0)} + \varepsilon\,\delta\!X + \mathcal{O}(\varepsilon^{2})$, where $X = g_{\alpha\beta}$, $A_{\alpha}$, $A^{'}_{\alpha}$, $v_{\alpha}$, $n_{\rm e}$, $P_{\rm t}$, $\Psi$, the superscript $(0)$ marks background quantities and $\varepsilon$ is a bookkeeping parameter. As we always consider a spherically symmetric background, we separate the angular dependence through a multipolar expansion.

Gravitational perturbations are expanded in tensor spherical harmonics. These fall into two classes, axial or polar, depending on their behaviour under parity transformations~\cite{Regge:1957td,Zerilli:1970se,Zerilli:1970wzz}. We can thus write
\begin{equation}
\delta g_{\alpha \beta}= \delta g_{\alpha \beta}^{\rm axial}+\delta g_{\alpha \beta}^{\rm polar}\,.
\end{equation}
In Regge-Wheeler gauge, this can be rewritten as
\begin{align}
\label{eq:Regge-Wheeler-gauge}
\renewcommand{\arraystretch}{1.25}
\setlength\arraycolsep{2pt}
\delta g_{\alpha \beta}\!=\!\sum_{\ell, m}\begin{pNiceArray}{cc|cc}[margin = 0.cm]
\Block[fill=blue!10,rounded-corners]{2-2}{}
H_0Y & H_1Y & \Block[fill=red!10,rounded-corners]{2-2}{}
        h_0 S_{\theta} & h_0 S_{\varphi} \\
H_1Y & H_2Y & h_1S_{\theta} & h_1 S_{\varphi} \\
\hline
\Block[fill=red!10,rounded-corners]{2-2}{}
* & * & \Block[fill=blue!10,rounded-corners]{2-2}{}
        r^2 KY & 0 \\
* & * & 0 & r^{2}\sin^{2}{\theta} KY \\
\end{pNiceArray}e^{-i \omega t}\,.
\end{align}
For convenience, we have denoted the axial $(\delta g_{\alpha \beta}^{\rm axial})$ and polar entries $(\delta g_{\alpha \beta}^{\rm polar})$ by red and blue, respectively. Furthermore, to avoid cluttering we leave out all dependencies on $(t,r,\theta,\varphi)$ as well as the $(\ell,m)$ label on the radial functions in the axial ($h^{\ell m}_0$ and $h^{\ell m}_1$) and polar ($H^{\ell m}_0$, $H^{\ell m}_1$, $H^{\ell m}_2$ and $K^{\ell m}$) sector, the spherical harmonics $Y = Y^{\ell m}$ and the axial vector harmonics $(S_{\theta}, S_{\varphi})=(S^{\ell m}_{\theta}, S^{\ell m}_{\varphi})  = (-\partial_{\varphi}Y^{\ell m}/\sin{\theta},\sin{\theta}\partial_{\theta}Y^{\ell m})$. Lastly, we take $\sum_{\ell,m} \equiv \sum_{\ell = 0}^{\infty}\sum_{m = -\ell}^{\ell}$. 

The EM field, DP field and the four-velocity can be expanded in a basis of vector spherical harmonics as
\begin{equation}
\label{eq:expansions}
\begin{aligned}
    \delta A_\alpha&=\frac{1}{r}\sum_{i=1}^4\sum_{\ell,m}c_iu_{i}^{\ell m}  Z_\alpha^{(i) \ell m} e^{-i \omega t}\,,\\
     \delta A'_\alpha&=\frac{1}{r}\sum_{i=1}^4\sum_{\ell,m}c_i {u'}_{i}^{\ell m}  Z_\alpha^{(i) \ell m} e^{-i \omega t}\,,\\
    \delta v_\alpha&=\frac{1}{r}\sum_{i=1}^4\sum_{\ell ,m}c_i v_{i}^{\ell m}  Z_\alpha^{(i)\ell m}e^{-i \omega t}\,,
\end{aligned}
\end{equation}
where $c_1=c_2=1$ and $c_3=c_4=1/\sqrt{\ell(\ell+1)}$ and the vector spherical harmonics
$Z_\alpha^{\ell m}$ are defined in e.g.~\cite{Rosa:2011my}. 

Finally, the density and pressure of the plasma and the axion field are expanded as
\begin{equation}
\begin{aligned}
\delta n_{\rm e} &= \sum_{\ell,m} n^{\ell m}_{\mathrm{e}} Y^{\ell m} e^{-i \omega t}\,,\\
\delta P_{\rm t} &= \sum_{\ell,m} P^{\ell m}_{\mathrm{t}} Y^{\ell m} e^{-i \omega t}\,, \\
\Psi &= \sum_{\ell,m} \psi^{\ell m} Y^{\ell m}e^{-i \omega t}\,,
\end{aligned}
\end{equation}
where we use scalar spherical harmonics.
%%%%%%%%%%%%%%%%%%%%%%%%
\section{Axion instability in an electric field}\label{appendix:flatinstab}
%%%%%%%%%%%%%%%%%%%%%%%%
In this appendix, we show how axionic instabilities can occur, even in flat spacetime. Consider $\sin\chi_0=0$ and a background ($\mathrm{B}$) system with a constant electric field along the $\hat{z}$ direction:
\begin{equation}\label{eq:constant_electric}
    A^{\rm B}_\mu=\left(z E_z,0,0,0\right)\,.
\end{equation}
We will also assume the presence of ions, which makes the plasma globally neutral. Under the influence of the electric field~\eqref{eq:constant_electric}, electrons are subject to a constant acceleration. From the momentum equation, one can easily infer the four velocity, given by
\begin{equation}
    v^{\rm B}_\mu=\left(\frac{\sqrt{m_{\rm e}^2+e^2 E_z^2 t^2}}{m_{\rm e}},0,0,\frac{e E_z t}{m_{\rm e}}\right) \,,
\end{equation}
such that $v^{\rm B}_\mu v^{\mathrm{B},\mu}=-1$. Likewise, ions feel a force in the opposite direction along the $\hat{z}$-axis. From the continuity equation, one can then infer the density profile of the fluid:
\begin{equation}\label{eq:density_profile}
    n^{\rm B}=\frac{n^{(0)}}{\sqrt{m_{\rm e}^2+ e^2 E_z^2 t^2}}\,,
\end{equation}
where $n^{(0)}$ is an integration constant. Given the time dependence of the four-velocity, the system is not stationary. To obtain a full solution, one should thus perform a consistent evolution in the time-domain. However, by assuming that the electric field is weak with respect to the particle's inertia, i.e., $e E_z\ll m_{\rm e}$, we can approximate the plasma as static on timescales $t\approx \mathcal{O}(m_{\rm e}/e E_z)$ and use standard frequency-domain methods. Moreover, in this approximation, the density is stationary and constant in space~\eqref{eq:density_profile}. Finally, as the ion's inertia is much larger than the electron's one, the same considerations hold.

We now perturb the fields in the frequency-domain as
\begin{equation}
\begin{aligned}
     \Psi & \sim \varepsilon \tilde{\psi} e^{-i (\omega t - k_i x^i)}\,, \\
    A_\mu & \sim A^{\rm B}_\mu+ \varepsilon \tilde{A}_\mu  e^{-i (\omega t - k_i x^i)}\,, \\ 
    v_\mu &\sim v^{\rm B}_\mu+ \varepsilon \tilde{v}_\mu  e^{-i (\omega t - k_i x^i)}\,, \\
    n_{\rm e} &\sim  n^{\rm B} + \varepsilon \tilde{n} e^{-i (\omega t - k_i x^i)} \,,   
\end{aligned}
\end{equation}
where $k_i$ is the wave vector in Fourier space and perturbed variables are marked with a tilde. Jointly solving the Maxwell and momentum equations in terms of $\tilde{\psi}$ and $\tilde{A}_z$ then yields
\begin{equation}\label{eq:flateqs}
\begin{aligned}
    \tilde{A}_0 &= -\frac{\tilde{A}_z \omega}{k_z}\,,\quad
    \tilde{A}_x = \frac{k_x \tilde{A}_z}{k_z}+\frac{2 i E_z k_y k_{\rm a} \tilde{\psi}}{\omega_{\rm p}^2+ k^2-\omega^2}\,, \\ 
    \tilde{A}_y &=  \frac{k_y \tilde{A}_z}{k_z}-\frac{2 i E_z k_x k_{\rm a} \tilde{\psi}}{\omega_{\rm p}^2+ k^2-\omega^2}\,, \\
    \tilde{v}_0 &= 0\,,\quad
    \tilde{v}_x = -\frac{2 i E_z k_y e k_{\rm a} \tilde{\psi}}{m_{\rm e}(\omega_{\rm p}^2+ k^2-\omega^2)}\,, \\
    \tilde{v}_y &= \frac{2 i E_z k_x e k_{\rm a} \tilde{\psi}}{m_{\rm e}(\omega_{\rm p}^2+ k^2-\omega^2)}\,, \quad
    \tilde{v}_z = 0\,,    
\end{aligned}
\end{equation}
where $k^2=k_i k^i$ and we defined the plasma frequency of the background plasma $\omega_{\rm p}^2=e^2 n^{\rm B}/m_{\rm e}$. Since $k_\mu \tilde{v}^\mu=0$, the EM perturbation is purely transverse. As a result, the continuity equation implies that density perturbations, which are longitudinal quantities, vanish identically, i.e., $\tilde{n}=0$. 

Finally, we plug these relations back in the Klein-Gordon equation. As the latter is sourced by the combination $k_y \tilde{A}_x-k_x \tilde{A}_y$, Eq.~\eqref{eq:flateqs} implies that $\tilde{A}_z$ vanishes, and we are left with a decoupled expression for $\tilde{\psi}$, which reads
\begin{equation}
 \left(-\omega^2+k^2+\mu_{\rm a}^2-\frac{4 E_z^2 (k_x^2+k_y^2)k_{\rm a}^2}{-\omega^2+k^2+\omega_{\rm p}^2} \right)\tilde{\psi}=0\,.
\end{equation}
We can then find the dispersion relation as
\begin{equation}
    \omega^2=\frac{1}{2}\left(2k^2+\mu_{\rm a}^2+\omega_{\rm p}^2 \pm \sqrt{(\mu_{\rm a}^2-\omega_{\rm p}^2)^2+16 E_z^2 k^2 k_{\rm a}^2}\right)\,.
\end{equation}
Clearly, the electric field has a critical threshold above which an instability arises. This threshold is given by
\begin{equation}
    E_z^{\rm crit}=\frac{\sqrt{k^2+\mu_{\rm a}^2}\sqrt{k^2+\omega_{\rm p}^2}}{2 k_{\rm a} k}\,,
\end{equation}
above which the system admits purely imaginary, unstable modes. For $\omega_{\rm p} \rightarrow 0$, one correctly obtains the vacuum threshold as found in~\cite{Boskovic:2018lkj}. Importantly, in the limit $\omega_{\rm p} \gg \mu_{\rm a}$, the threshold increases, as the axion-photon conversion angle drops. 

\begin{figure}
    \centering
    \includegraphics[width = \linewidth]{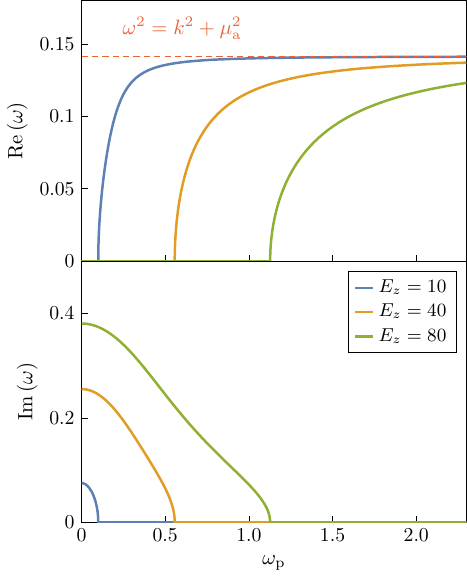}
    \caption{Spectrum of the axionic modes as a function of the plasma frequency for three different choices of the electric field strength. Increasing the plasma frequency stabilizes the system, as modes turn from imaginary to real, and the frequency approaches the axion vacuum dispersion relation. The parameters in this plot are $k_{\rm a}=0.01$, $k^2=0.1$ and $\mu_{\rm a}=0.1$ in arbitrary units.}
    \label{fig:flatmodes}
\end{figure}

Figure~\ref{fig:flatmodes} shows the real and imaginary part of the axionic frequency as a function of the plasma frequency for different values of the electric field. In all three cases, the electric field is above the threshold value for low plasma frequencies, and the modes are purely imaginary. By increasing the plasma frequency, the system stabilizes as the modes become purely real while they approach the vacuum dispersion relation of the axion. This is exactly the impact of in-medium suppression, as in the limit $\omega_{\rm p} \gg \mu_{\rm a}$ the axion effectively decouples from the photon. For small electric fields (blue line), this happens precisely at the threshold $\omega_{\rm p}=\mu_{\rm a}$.

%
%%%%%%%%%%%%%%%%%%%%%%%%
\section{Dark photon basis}\label{appendix:DPbasis}
%%%%%%%%%%%%%%%%%%%%%%%%
\subsection{Basis choices}
%%%%%%%%%%%%%%%%%%%%%%%%
Kinetic mixing between the Standard Model and dark photon can be described by the Lagrangian 
\begin{equation}
    \mathcal{L}=\mathcal{L}_{\rm EM}+\mathcal{L}_{\rm{Proca}}+\frac{1}{2}\sin \chi_0 F'_{\mu\nu}F^{\mu\nu}\,.
\end{equation}
To make the equations more tractable, the mixing term can be removed through a field redefinition. Two different choices are possible. The redefinition $A'_\mu \rightarrow A'_\mu+\sin\chi_0 A_\mu$ leads to the ``interaction basis'', with a Lagrangian
\begin{equation}\label{eq:lagrangian_interaction}
    \begin{aligned}
        &\mathcal{L_{\rm inter}}=-\frac{1}{4}\left(F_{\mu\nu}F^{\mu\nu}+F'_{\mu\nu}F'^{ \mu\nu}\right)-\frac{\mu_{\gamma '}^2}{2} A'^\mu A'_\mu \\&- \mu_{\gamma'}^2\sin\chi_0 A'_\mu A^\mu+ j^\mu A_\mu \,.
\end{aligned}
\end{equation}
In this basis, the fields are directly coupled, as clearly shown in~\eqref{eq:evoleqns}, while only the visible field $A_\mu$ couples with the Standard Model current $j_\mu$. This is the basis we use in the main text.

An alternative choice which removes the kinetic mixing is $A_\mu \rightarrow A_\mu+\sin\chi_0 A'_\mu$, which leads to the ``mass basis'' with a Lagrangian:
\begin{equation}\label{eq:lagrangian_mass}
    \begin{aligned}
        &\mathcal{L_{\rm mass}}=-\frac{1}{4}\left(F_{\mu\nu}F^{\mu\nu}+F'_{\mu\nu}F'^{ \mu\nu}\right)-\frac{\mu_{\gamma '}^2}{2} A'^\mu A'_\mu \\&+ j^\mu (A_\mu+ \sin \chi_0 A'_\mu) \,.
\end{aligned}
\end{equation}
The corresponding field equations are given by
\begin{equation}\label{eq:masseqs}
\begin{aligned}
    \nabla_\nu F^{\mu \nu} &=j^\mu\,, \\
    \nabla_\nu F'^{\mu \nu} &=-\mu_{\gamma'}^2 A'^\mu+ \sin \chi_0 j^\mu\,,
\end{aligned}
\end{equation}
while the momentum equation in the Einstein cluster setup reads
\begin{equation}
    \nabla^\nu T^{\rm p}_{\mu\nu}=e n_{\rm e} (F_{\mu\nu}+ \sin \chi_0 F'_{\mu\nu})\,.
\end{equation}
In this case, the two fields are not directly coupled, yet both of them are coupled to the electrons. Note that, in this basis $A^\mu$ is not the visible photon, i.e., the one accelerating charged particles. Instead, the latter is given by the combination $A_{\rm{obs}}=A_\mu+\sin\chi_0 A'_\mu$. 

As it should, the physics in the two bases is equivalent, and one can simply choose a preferred basis depending on the problem at hand. 
%%%%%%%%%%%%%%%%%%%%%%%%
\subsection{Black hole perturbation theory in the mass basis}
%%%%%%%%%%%%%%%%%%%%%%%%
For completeness, we briefly outline the computations performed in the main text, yet now in the mass basis. This also allows for a more easy comparison with previous work~\cite{Caputo:2021efm}. Performing the multipolar decomposition of the momentum equation yields the following relation between the axial four-velocity and the EM and DP axial fields:
\begin{equation}\label{eq:DP_massbasis}
v_{4}=-\frac{e}{m_{\rm e}}(u^{\rm b}_{4}+ \sin \chi_0 u_4^{\rm b} {}')\,,
\end{equation}
where we use the superscript $\mathrm{b}$ to refer to mass basis quantities. From Eq.~\eqref{eq:DP_massbasis}, one can immediately see that in this basis, the DP field does influences the motion of charged particles. One can then expand the equations~\eqref{eq:masseqs} to find a set of coupled partial differential equations:
\begin{equation}
\label{eq:wavelike-eqn_DP_mass}
\begin{aligned}
\hat{\mathcal{L}}u^{\rm b}_{4}&=f\left(\omega_{\rm p}^2 + \frac{\lambda}{r^{2}}\right) u^{\rm b}_{4}+f \omega_{\rm p}^2 \sin{\chi_0}\,u^{\rm b}_{4} {}'\,,
 \\
\hat{\mathcal{L}}u^{\rm b}_{4} {}'&=f\left(\mu_{\gamma'}^2 + \frac{\lambda}{r^{2}}\right) u^{\rm b}_{4} {}'+f \omega_{\rm p}^2 \sin{\chi_0}\,u^{\rm b}_4\,.
\end{aligned}
\end{equation}
The visible photon is then given by $u_4=u^{\rm b}_4+ \sin \chi_0 u^{\rm b}_4 {}'$. One can thus do the same computations as the main text in the mass basis, and compare the results through this linear combination.
\clearpage
\bibliography{ref}
\end{document}